\documentclass[10pt,pra,showpacs,twocolumn,superscriptaddress]{revtex4}%
\usepackage{amsmath}
\usepackage{amsfonts}
\usepackage{amssymb}
\usepackage{graphicx}
\begin{document}
\title{Two-way Gaussian quantum cryptography against coherent attacks in direct reconciliation}

\begin{abstract}
We consider the two-way quantum cryptographic protocol with coherent states
assuming direct reconciliation. A detailed security analysis is performed
considering a two-mode coherent attack, that represents the residual
eavesdropping once the parties have reduced the general attack by applying
symmetric random permutations. In this context we provide a general analytical
expression for the keyrate, discussing the impact of the residual two-mode
correlations on the security of the scheme. In particular we identify the
optimal eavesdropping against two-way quantum communication, which is given by a two-mode coherent attack with
symmetric and separable correlations.

\end{abstract}

\author{Carlo Ottaviani}
\email{Carlo.Ottaviani@york.ac.uk}
\affiliation{Department of Computer Science \& York Center for Quantum Technologies,
University of York, York YO10 5GH, United Kingdom}
\author{Stefano Mancini}
\affiliation{School of Science and Technology, University of Camerino, 62032 Camerino,
Italy \& INFN Sezione di Perugia 06123 Perugia, Italy}
\author{Stefano Pirandola}
\affiliation{Department of Computer Science \& York Center for Quantum Technologies,
University of York, York YO10 5GH, United Kingdom}
\maketitle

\section{Introduction}

The goal of quantum key distribution (QKD) \cite{SCARANI} is to make available
unconditionally secure private keys between two authenticated users, Alice and
Bob. Carriers of the information are quantum systems whose quantum nature is
exploited to generate the same random sequence of bits, to be then used as a
cryptographic key in one-time pad protocols. This strategy is based on the
fundamental restriction, imposed by quantum mechanics, that obtaining perfect
copies of arbitrary quantum states is impossible. In fact, any attempt in this
sense unavoidably introduces some noise perturbing the quantum state itself
(no-cloning theorem \cite{NOCLONING}).

To convert this feature of the quantum world into the ultimate cipher
\cite{BB84}, any quantum cryptographic protocol needs to be arranged in a
first quantum communication step, followed by a classical communication one.
During the first stage, Alice encodes classical information into
non-orthogonal quantum states, which are sent to Bob over a noisy quantum
channel. This is used $N$ times and assumed to be in the hands of an
eavesdropper (Eve). The quantum signals are measured by Bob, detecting a noisy
version of Alice's quantum states. After many uses of the channel ($N>>1$),
the parties can share a random sequence of bits called the raw key. At this
point, the parties sacrifice part of the $N$ bits, from the raw key,
communicating over a classical public channel. This allows them to compare the
data in their hands, and to estimate the presence of the eavesdropper on the
quantum channel. This second stage allows Alice and Bob to quantify the
adequate amount of error correction and privacy amplifications needed to
reduce the stolen information to a negligible amount \cite{PRIVACY}.

In recent years, continuous variable (CV) quantum systems
\cite{RMP-SAM,RMP-GAUSS} have attracted increasing attention for the
implementation of quantum communication tasks, with special attention devoted
to Gaussian CV states. The appealing possibilities of this approach are based
on the replacement of single photon pulses with bright coherent states, and
single photon detection with simpler and more efficient Gaussian operations
like homodyne and/or heterodyne detection schemes. This simplifies the
experimental realization on one side, and can increases the key-rate
production of the protocols by many orders of magnitude on the other
\cite{SCARANI,RELAY,SIMMETRICO,MDI-CORR,GAE}. Furthermore, Gaussian CV
protocols can easily go broadband. Within this research area, quantum
cryptography has been one of the most prolific field of the last years
\cite{RMP-GAUSS}, with extensive theoretical and experimental research
developed to improve the performances of point-to-point communications in
one-way \cite{GROSS-NAT,NOSWITCHING} and two-way \cite{PIRS2,WEED-th2} protocols.

In two-way schemes the parties exploit twice the quantum channel per each use
of the protocol \cite{PIRS2,WEED-th2} (see also Ref.~\cite{MARCO} for DV
two-way protocols and Ref.~\cite{SHAPIRO1,SHAPIRO2,SHAPIRO3} for CV two-way
protocols based on quantum illumination \cite{refQI1,refQI2,refQI3}). In particular CV two-way
protocols \cite{PIRS2,WEED-th2} can achieve higher security thresholds thanks
to an improved tolerance to the eavesdropper's noise. In fact, the analysis
developed in Ref.~\cite{PIRS2} (see for example Fig.$~3$ of Ref.~\cite{PIRS2})
proved that, for fixed values of channel's transmissivity, CV two-way
protocols tolerate higher level of noise than one-way in the presence of
collective attacks. This makes this approach appealing to achieve high-rate
secure communication in noisier environments, where one-way communication
fails to provide a secure key.

\begin{figure}[t]
\vspace{-0.0cm}
\par
\begin{center}
\includegraphics[width=0.45\textwidth]{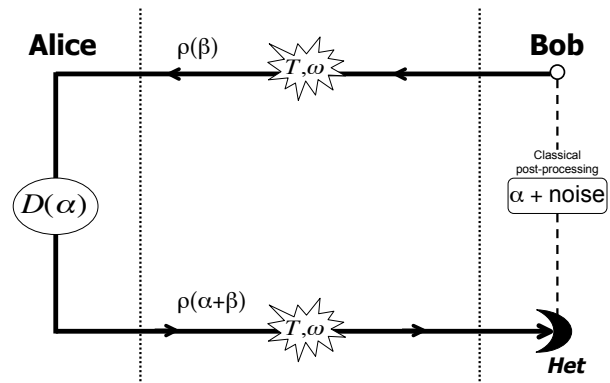}
\end{center}
\par
\vspace{-0.0cm} \caption{(Color online)In the general two-way protocol Bob
sends the reference state $\rho(\beta)$ to Alice who applies a random
displacement $D(\alpha)$. The resulting Gaussian state $\rho(\alpha+\beta)$ is
sent back to Bob who applies heterodyne detection and classical
post-processing to recover Alice's encoding $(\alpha)$.}%
\label{schemePM}%
\end{figure}

In this work we study the security of two-way QKD considering general coherent
attacks and focusing on direct reconciliation. In this case, see
Fig.~\ref{schemePM}, Gaussian-modulated reference coherent states
$|\beta\rangle$, are sent from Bob to Alice through the quantum channel, and
are processed by Alice via a random displacement operation $D(\alpha)$, with
Gaussian modulation of amplitudes $\alpha$. The output $\rho(\alpha,\beta)$ is
sent backward to Bob who applies heterodyne detection and classical
post-processing, in order to subtract the reference amplitude $\beta$ and
infer Alice's signal amplitude $\alpha$. The higher tolerance to noise,
granted by the double use of the quantum channel, is due to the fact that Eve
needs to attack both the forward and backward steps of the quantum
communication, in order\ to extract information on both $\beta$ and $\alpha$
\cite{PIRS2}.

The key-rate of the two-way QKD protocol has been studied under the standard
assumption of collective Gaussian attacks \cite{RMP-GAUSS}. Protection against
coherent attacks can be achieved switching randomly between the single and
double use of the quantum channel (ON/OFF switching) \cite{PIRS2}. Collective
attacks means that Eve attaches uncorrelated ancillary modes to each use of
the quantum channel. The ancillas interact unitarily with the communication
modes and are then measured by the eavesdropper. In this scenario, recently,
it has been possible to extend two-way QKD also to the case where the parties
encode information affected by trusted thermal noise \cite{WEED-th2}.

In the present study we explicitly derive the secret-key of the two-way
protocol in the case where Eve's ancillary states are correlated. In such a
case the Alice-Bob communication line becomes a memory channel \cite{PIRS-NJP,
LUPO} in contrast to the case of collective attacks where it is memoryless.
Ours is the first security analysis of a two-way CV-QKD protocol against
coherent attacks. Our analysis is based on the conventional assumption that
the parties exchange a large number of signals ($N>>1$). In this case we can
reduce the general attack to a simpler two-mode coherent attack where, for
each use of the protocol, Eve's ancillas share non-zero two-mode correlations.
In addition to that we also consider the case of asymptotically large Gaussian
modulation of the amplitudes $\alpha$ and $\beta$. This allows us to work with
analytical mathematical expressions, and to find the optimal two-mode coherent
attack against the protocol, when Eve injects symmetric separable correlations
\cite{PIRS-NJP}.

The results for the two-way protocol are compared with the performances of the
one-way version of the scheme, and show that eavesdropping two-way quantum
communication with a suitable two-mode coherent attack can reduce the
performances partly below the one-way security threshold. This represents the
first example of a coherent attack overcoming the performances of collective
ones, in point-to-point protocols. We discuss why this happens, in the context
considered here, and finally we compare our results with other recent studies
\cite{RELAY, SIMMETRICO,2WAY-carlo} where two-mode optimal coherent attacks
have been identified for end-to-end cryptographic protocols.

Our results are important for the development of the security analysis of
continuous variable protocols, and to identify the general challenges to
implement secure point-to-point communications. Our results confirm that the
ON/OFF switching operated by Alice, described in detail in \ Refs.
\cite{PIRS2} and \cite{2WAY-carlo}, represents a necessary countermeasure to
overcome the problem of realistic coherent attacks in two-way point-to-point
quantum cryptography.

The structure of this paper is the following. In Sec. \ref{protocol-crypto} we
introduce the protocol and illustrate the reduction of the general
eavesdropping to a two-mode coherent attack. In Sec. \ref{RATE} we provide the
definition of the key-rate and we show how to compute the Holevo bound and
Alice-Bob mutual information, arriving at the analytical expression of the
secret-key rate. In Sec. \ref{results} we analyze the security thresholds and
we study the behavior of the relevant quantities as function of Eve's injected
thermal noise and degree of two-mode correlation. Sec. \ref{conclusion} is for conclusions.

\section{Protocol and eavesdropping\label{protocol-crypto}}

We show the protocol in the entanglement based representation (see Fig
\ref{scheme}). We reduce the general coherent eavesdropping to two-mode
coherent attacks, and we illustrate the steps to compute the total and
conditional covariance matrices. Then in the next section we provide the
analytical expression of the symplectic spectra which are used to compute the
Holevo bound.

\subsection{Coherent Gaussian Attack}

In a general (coherent) eavesdropping, Eve processes all the $N$ uses of the
quantum channel applying a global coherent unitary operation that correlates
all the modes involved in the different uses. However, exploiting the quantum
de Finetti theorem \cite{deFINETTI} for infinite-dimensional systems, this
general scenario can be reduced to a two-mode coherent attack. The parties can
apply symmetric random permutations of the classical data in such a way that
for $N>>1$, the cross correlations between distinct uses of the two-way
communication can be neglected. The global coherence of the attack is so
reduced to a two-mode coherence, between the forward and backward channels
involved\ in each round-trip quantum communication.

This residual two-mode coherent attack, in the most typical case, is
implemented by two beam splitters of transmissivity $T$ \cite{PIRS-ATTACKS},
where Eve mixes two ancillary modes $E_{1}$ and $E_{2}$ (see Fig.
\ref{scheme}). These two ancillas belong to generally larger set of modes,
$\{E_{1}$, $E_{2},\mathbf{e}\}$, defining the pure initial quantum state owned
by the eavesdropper. The two-mode Gaussian state $\rho_{E_{1}E_{2}}$ is
generally-correlated and described by the following covariance matrix (CM)%
\begin{equation}
\mathbf{V}_{E_{1}E_{2}}=\left(
\begin{array}
[c]{cc}%
\omega\mathbf{I} & \mathbf{G}\\
\mathbf{G} & \omega\mathbf{I}%
\end{array}
\right)  ,\text{ }\mathbf{G:=}\left(
\begin{array}
[c]{cc}%
g & 0\\
0 & g^{\prime}%
\end{array}
\right)  . \label{VE1E2}%
\end{equation}
Here the parameter $\omega$ describes the variance of the thermal noise
injected by Eve in the beam splitters, $\mathbf{I}=$ diag($1,1$),
$\mathbf{Z}=$ diag($1,-1$), and matrix $\mathbf{G}$ accounts for the specific
two-mode correlations employed by Eve to eavesdrop. The parameters $\omega,$
$g$ and $g^{\prime}$ must fulfill the conditions given in Ref. \cite{PIRS-NJP}%
, in order to represent a physical attack. Note that the properties of this
type of non-Markovian channel have been recently studied in the context of
relay-based continuous variable quantum cryptography \cite{RELAY,SIMMETRICO},
where they have been also classified and grouped in three possible cases. More
recently it has been shown how they could be exploited to reactivate
entanglement distribution and quantum communication protocols
\cite{REACTIVATION}.

We distinguish between three possible extremal cases: \textit{Collective
attacks} for $g=g^{\prime}=0$ corresponding to the standard collective
eavesdropping; \textit{separable attacks} defined by the condition
$|g|=|g^{\prime}|=\omega-1$, representing coherent attacks with separable
correlations injected and, finally, Einstein-Podolsky-Rosen (\textit{EPR attacks}) where $g=-g^{\prime
}=\sqrt{\omega^{2}-1}$ and $g=-g^{\prime}=-\sqrt{\omega^{2}-1}$. These three
eavesdropping strategies are not equivalent, and in next section we will
identify the optimal one.

\subsection{Entanglement based protocol\label{sectionEB}}

\begin{figure}[ptb]
\vspace{-0.0cm}
\par
\begin{center}
\includegraphics[width=0.45\textwidth]{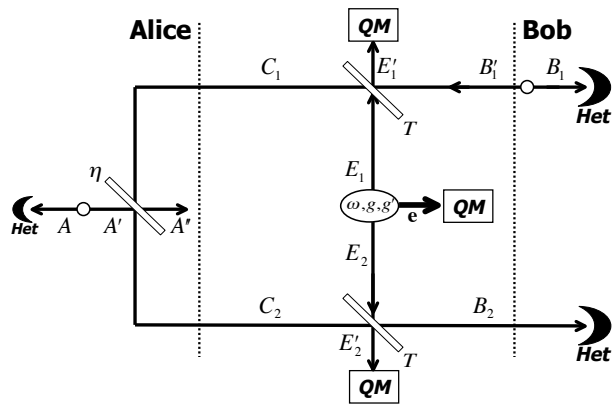}
\end{center}
\par
\vspace{-0.0cm}\caption{(Color online)Entanglement based representation of the
two-way QKD protocol. Bob prepares reference coherent states $|\beta\rangle$.
This can be done by heterodyning one part of an EPR state. One mode is
measured $(B_{1})$, while the other, $B_{1}^{\prime}$ is sent to Alice through
an insecure quantum channel. Alice applies a random displacement of the
reference state, $D(\alpha)$, which can be implemented by a beam splitter with
transmissivity $\eta$ and another EPR state. Choosing appropriately the
transmissivity $\eta$, and the variance of her EPR state, Alice sends
displaced output state $\rho(\alpha,\beta)$ back to Bob. These are heterodyned
and classically post-processed by Bob. In this way he recovers Alice's
encoding by subtracting the known reference amplitude $\beta$. The information
encoded in the amplitude $\alpha$ is then used to obtain the raw key.}%
\label{scheme}%
\end{figure}

We perform the security analysis in the entanglement based representation so
that, besides previous dilation of the quantum channel, we also provide the
purification of the source of Bob's coherent states and Alice's random
displacements. Thus, by referring to Fig. \ref{scheme}, we first assume that
Bob's coherent states originate from two-mode squeezed vacuum states (EPR
states), which are zero mean Gaussian state is described by the CM%
\begin{equation}
\mathbf{V}_{B_{1}B_{1}^{\prime}}=\left(
\begin{array}
[c]{cc}%
\mu_{B}\mathbf{I} & \sqrt{\mu_{B}^{2}-1}\mathbf{Z}\\
\sqrt{\mu_{B}^{2}-1}\mathbf{Z} & \mu_{B}\mathbf{I}%
\end{array}
\right)  . \label{VBob}%
\end{equation}
where the variance parameter $\mu_{B}$ quantifies the entanglement and also
the local thermal noise in modes $B_{1}$ and $B_{1}^{\prime}$. The heterodyne
measurement performed by Bob on mode $B_{1}$, remotely projects mode
$B_{1}^{\prime}$ on a coherent state traveling forward (from Bob to Alice)
through the quantum channel. Its amplitude is classically modulated with a
Gaussian distribution having variance $\mu=\mu_{B}-1$.

At Alice's station the random displacement $D(\alpha)$ can be implemented by
means of a beam splitter of transmissivity $\eta$. This mixes the incoming
mode $C_{1}$ with a mode $A^{\prime}$, coming from Alice's EPR pairs $A$ and
$A^{\prime}$, whose Gaussian quantum state, $\rho_{AA^{\prime}}$, is described
by the following CM%
\begin{equation}
\mathbf{V}_{AA^{\prime}}=\left(
\begin{array}
[c]{cc}%
\mu_{A}\mathbf{I} & \sqrt{\mu_{A}^{2}-1}\mathbf{Z}\\
\sqrt{\mu_{A}^{2}-1}\mathbf{Z} & \mu_{A}\mathbf{I}%
\end{array}
\right)  .
\end{equation}
While Alice's mode $A^{\prime}$ is sent through the beam splitter, the other
mode $A$ is heterodyne detected, in order to project the mode $A^{\prime}$
onto a coherent state $|\gamma\rangle$ modulated with variance $\mu_{\gamma}$
such that%
\begin{equation}
\mu_{\gamma}=\mu_{A}-1.
\end{equation}
This setup is a way to equivalently simulate Alice's random displacements. In
fact, for simplicity, consider the case where Eve is absent and there is no
loss and noise in the quantum channel. In this scenario Alice receives
$|\beta\rangle$ and must send $|\beta+\alpha\rangle=D(\alpha)|\beta\rangle$
back to Bob. We can see that using the setup with the beam splitter, Alice
prepares her output mode $C_{2}$ into the coherent state%
\begin{equation}
|\sqrt{\eta}\beta+\sqrt{1-\eta}\gamma\rangle. \label{COHC2}%
\end{equation}
Now, in order to obtain a coherent state of the form $|\beta+\alpha\rangle$
from Eq. (\ref{COHC2}), we design Alice's beam splitter to have transmissivity
$\eta\rightarrow1$, and we assume that the coherent amplitude $\gamma
\rightarrow\infty$ in such way that%
\[
\gamma=\frac{\alpha}{\sqrt{1-\eta}}.
\]
This is possible in theory by using an EPR input state for Alice with
divergent variance $\mu_{\gamma}+1$ where
\begin{equation}
\mu_{\gamma}:=\frac{\mu}{1-\eta}.
\end{equation}
Under these assumptions we get%
\[
|\sqrt{\eta}\beta+\sqrt{1-\eta}\gamma\rangle\simeq|\beta+\alpha\rangle.
\]

\section{Key-rate, Holevo function and mutual information\label{RATE}}

In direct reconciliation the parties use Alice's amplitudes $\alpha$ to
prepare the secret key. This means that, during the classical procedure of
parameter estimation, error correction and privacy amplification, Bob infers
the values of Alice's variables $\alpha$ from the results of his measurements.
The security performances are quantified by the asymptotic secret-key rate%
\begin{equation}
R:=I_{AB}-\chi_{EA} \label{RATE-GEN}%
\end{equation}
which is defined as the difference between Alice-Bob's mutual information
$I_{AB}$ and the Holevo function $\chi_{EA}$ which upper bounds Eve-Alice's
mutual information.

The advantage of using the entanglement based representation od Sec.
\ref{sectionEB}, relies on the fact that we do not need to know the details of
the coherent operations performed by Eve on the modes. Instead, we can compute
the function $\chi_{EA}$ from the output quantum state of Alice and Bob
\cite{RMP-GAUSS}. More precisely we compute Eve's Holevo information as%
\begin{equation}
\chi_{EA}=S_{E}-S_{E|\alpha}, \label{HOLEVO}%
\end{equation}
where $S_{E}$ is the von Neumann entropy of Eve's total output modes, which
coincides with the von Neumann of Alice's and Bob's total output modes $B_{1}%
$, $A$,$A^{\prime\prime}$, $B_{2}$. The other quantity is the von Neumann
entropy of Eve's output modes conditioned on Alice's detection $\alpha$. This
is equal to the von Neumann entropy of Bob's output modes $B_{1}$, $B_{2}$
conditioned on $\alpha$.

For Gaussian states, the von Neumann entropy has a particularly simple form in
terms of the symplectic eigenvalues \cite{RMP-GAUSS}. It is given by%
\begin{equation}
S:=\sum_{\nu}h(\nu), \label{SGEN}%
\end{equation}
where $\nu$ are the symplectic eigenvalues of the CM associated with the
state, and the entropic function $h(\nu)$ is defined as%
\[
h(\nu):=\frac{\nu+1}{2}\log_{2}\frac{\nu+1}{2}-\frac{\nu-1}{2}\log_{2}%
\frac{\nu-1}{2}.
\]
This expression simplifies further in the limit of large modulation $\mu>>1$,
in which case we have%
\begin{equation}
h(\nu)\rightarrow\log_{2}\frac{e}{2}\nu+O(\nu^{-1}). \label{gASY}%
\end{equation}

In next subsection we provide the total and conditional CMs corresponding to
$\rho_{B_{1}AA^{\prime\prime}B_{2}}$ and $\rho_{B_{1}B_{2}|\alpha}$ and the
respective symplectic spectra, that are then used to compute the Holevo bound
$\chi_{EA}$.

\subsection{Total symplectic spectrum}

The global Alice-Bob quantum state, $\rho_{B_{1}AA^{\prime\prime}B_{2}}$, is a
Gaussian state whose properties are described by the following CM (we use the
modes ordering $B_{1}AA^{\prime\prime}B_{2}$)%
\begin{equation}
\mathbf{V}=\left(
\begin{array}
[c]{cccc}%
\mu_{B}\mathbf{I} &  & \phi\mathbf{Z} & \theta\mathbf{Z}\\
& \mu_{A}\mathbf{I} & \xi\mathbf{Z} & \tau\mathbf{Z}\\
\phi\mathbf{Z} & \xi\mathbf{Z} & k\mathbf{I} & \delta\mathbf{I}\\
\theta\mathbf{Z} & \tau\mathbf{Z} & \delta\mathbf{I} & \varepsilon\mathbf{I}%
\end{array}
\right)  +\left(
\begin{array}
[c]{cccc}
&  &  & \\
&  &  & \\
&  &  & g_{\delta}\mathbf{G}\\
&  & g_{\delta}\mathbf{G} & g_{\varepsilon}\mathbf{G}%
\end{array}
\right)  , \label{VTOT}%
\end{equation}
where the missing matrix entries are zero and we have defined%
\begin{align}
\phi &  :=-\sqrt{T(1-\eta)(\mu_{B}^{2}-1)},\nonumber\\
\theta &  :=T\sqrt{\eta(\mu_{B}^{2}-1)},\nonumber\\
k  &  :=\eta\mu_{A}+(1-\eta)[T\mu_{B}+(1-T)\omega],\nonumber\\
\xi &  :=\sqrt{\eta(\mu_{A}^{2}-1)},\nonumber\\
\tau &  :=\sqrt{T(1-\eta)(\mu_{A}^{2}-1)},\nonumber\\
\varepsilon &  :=T^{2}\eta\mu_{B}+T(1-\eta)\mu_{A}+(T\eta+1)(1-T)\omega
,\nonumber\\
g_{\varepsilon}  &  :=2(1-T)\sqrt{\eta T},\nonumber\\
\delta &  :=\sqrt{T\eta(1-\eta)}[\mu_{A}-T\mu_{B}-(1-T)\omega],\nonumber\\
g_{\delta}  &  :=-(1-T)\sqrt{(1-\eta)}.
\end{align}

To obtain the symplectic spectrum of the CM of Eq. (\ref{VTOT}), we first
compute the matrix%
\begin{equation}
\mathbf{M}_{T}=i\mathbf{\Omega V}, \label{MTOT}%
\end{equation}
where $\mathbf{\Omega=}\bigoplus_{k=1}^{4}\mathbf{\tilde{\omega}}_{k}$, with
$\mathbf{\tilde{\omega}}_{k}=\left(
\begin{array}
[c]{cc}%
0 & 1\\
-1 & 0
\end{array}
\right)  $ is the symplectic form. Then we compute the standard eigenvalues of
Eq. (\ref{MTOT}). After simple algebra and taking the limit of large
modulation ($\mu>>1$), we find the following general expressions
\begin{align}
\nu_{1}  &  =\sqrt{(\omega-g)(\omega-g^{\prime})},\label{NI1}\\
\nu_{2}  &  =\sqrt{(\omega+g)(\omega+g^{\prime})},\label{NI2}\\
\nu_{3}\nu_{4}  &  =(1-T)^{2}\mu^{2},
\end{align}
where the dependency on the correlation parameters, $g$ and $g^{\prime}$,
generalizes the known total symplectic spectrum under collective attacks
\cite{PIRS2}, recovered for $g=g^{\prime}=0$. Using this spectrum with Eqs.
(\ref{SGEN}) and (\ref{gASY}), one easily obtains the asymptotic total von
Neumann entropy, that we can write as%
\begin{equation}
S_{E}=h(\nu_{1})+h(\nu_{2})+\log_{2}\frac{e^{2}}{4}(1-T)^{2}\mu^{2}.
\label{SE}%
\end{equation}

\subsection{Conditional symplectic spectrum and Holevo bound}

When the protocol is used in direct reconciliation Bob's conditional CM can be
obtained straightforwardly considering the CM involving Bob's modes, obtained
from Eq. (\ref{VTOT}) tracing out Alice's modes. This approach considerably
simplifies the problem. Starting from the following matrix%
\begin{equation}
\mathbf{V}_{B_{1}B_{2}}=\left(
\begin{array}
[c]{cc}%
\mu_{B}\mathbf{I} & \theta\mathbf{Z}\\
\theta\mathbf{Z} & \varepsilon\mathbf{I}+g_{\varepsilon}\mathbf{G}%
\end{array}
\right)  , \label{VBOB}%
\end{equation}
we set $\mu_{A}=1$ to simulate the conditioning on Alice's measurements, to
arrive at the conditional CM given by%
\begin{equation}
\mathbf{V}_{C}=\mathbf{V}_{B_{1}B_{2}}(\mu_{A}=1). \label{VC}%
\end{equation}

From this CM we compute the matrix%
\begin{equation}
\mathbf{M}_{C}=i\mathbf{\Omega V}_{C},
\end{equation}
where $\mathbf{\Omega=}\bigoplus_{k=1}^{2}\mathbf{\tilde{\omega}}_{k}$, and we
derive its spectrum. Considering the asymptotic limit for large $\mu$ and the
limit $\eta\rightarrow1$, we obtain the following pair of symplectic
eigenvalues%
\begin{align}
\bar{\nu}_{1}  &  =\sqrt{\omega+2g\frac{\sqrt{T}}{1+T}}\sqrt{\omega
+2g^{\prime}\frac{\sqrt{T}}{1+T}},\label{NIBAR1}\\
\bar{\nu}_{2}  &  =(1-T^{2})\mu.\nonumber
\end{align}
Using $\bar{\nu}_{1}$ and $\bar{\nu}_{2}$ in Eq. (\ref{SGEN}) and
(\ref{gASY}), we derive the conditional von Neumann entropy%
\begin{align}
S_{E|\alpha}  &  =h(\bar{\nu}_{1})+h(\bar{\nu}_{2}),\nonumber\\
&  =h(\bar{\nu}_{1})+\log_{2}\frac{e}{2}(1-T^{2})\mu. \label{SEC}%
\end{align}
Finally, putting together the results of Eqs. (\ref{SE}) and (\ref{SEC}) in
the definition of the Holevo function, Eq. (\ref{HOLEVO}), we find the
analytic expression of the Holevo bound%
\begin{equation}
\chi_{EA}=h(\nu_{1})+h(\nu_{2})-h(\bar{\nu}_{1})+\log_{2}\frac{e}{2}\frac
{1-T}{1+T}\mu. \label{CHI}%
\end{equation}

\subsection{Mutual Information}

To obtain the secret-key rate we also need Alice-Bob mutual information. Since
both quadratures, $q$ and $p$, of mode $B_{2}$ are measured, the mutual
information $I_{AB}$ is given by the following expression%
\[
I_{AB}=\frac{1}{2}\log_{2}\frac{V_{B}^{q}+1}{V_{B|\alpha\beta}^{q}+1}+\frac
{1}{2}\log_{2}\frac{V_{B}^{p}+1}{V_{B|\alpha\beta}^{p}+1},
\]
where $V_{B}^{q}$, $V_{B}^{p}$ represent the variances for quadratures $q$ and
$p$ of mode $B_{2}$, while $V_{B|\alpha\beta}^{q}$ and $V_{B|\alpha\beta}^{p}$
describe the conditional variances after Bob and Alice's measurements. The
former can be obtained from the diagonal block of the CM given in Eq.
(\ref{VBOB}), describing mode $B_{2}$. This is given by the expression
\begin{equation}
\mathbf{B}_{2}=\varepsilon\mathbf{I}+g_{\varepsilon}\mathbf{G,}%
\end{equation}
from which, taking the limit $\eta\rightarrow1$ and setting $\mu_{B}=1$, we
obtain
\begin{align*}
V_{B}^{q}  &  =T^{2}+T\mu+(1-T^{2})\omega+2g(1-T)\sqrt{T},\\
V_{B}^{p}  &  =T^{2}+T\mu+(1-T^{2})\omega+2g^{\prime}(1-T)\sqrt{T}.
\end{align*}

The conditional variances, $V_{B|\alpha\beta}^{q}$ and $V_{B|\alpha\beta}^{p}%
$, can now be obtained setting $\mu=0$ in the previous equations. Taking the
limit of large modulation $\mu>>1$, we get the asymptotic Alice-Bob mutual
information
\begin{equation}
I_{AB}=\frac{1}{2}\log_{2}\frac{T^{2}\mu^{2}}{\sigma\sigma^{\prime}},
\label{IAB-ASY}%
\end{equation}
where%
\begin{align*}
\sigma &  :=V_{B|\alpha\beta}^{q}+1=\Delta+2g(1-T)\sqrt{T},\\
\sigma^{\prime}  &  :=V_{B|\alpha\beta}^{p}+1=\Delta+2g^{\prime}(1-T)\sqrt{T},
\end{align*}
and%
\[
\Delta:=1+T^{2}+(1-T^{2})\omega.
\]

\subsection{Secret-key rate}

We have now all the quantities needed to compute the secret-key rate defined
in Eq. (\ref{RATE-GEN}). From the expressions for the asymptotic mutual
information given in Eq. (\ref{IAB-ASY}), and the Holevo bound of Eq.
(\ref{CHI}), after some simple algebra we get the following formula for the
key-rate
\begin{equation}
R=\log_{2}\frac{2T(1+T)}{e(1-T)\sqrt{\sigma\sigma^{\prime}}}-h(\nu_{1}%
)-h(\nu_{2})+h(\bar{\nu}_{1}), \label{KEYRATE-AS}%
\end{equation}
where $\nu_{1}$ and $\nu_{2}$ are given in Eq. (\ref{NI1}) and (\ref{NI2}) and
$\bar{\nu}_{1}$ is given in Eq. (\ref{NIBAR1}).

\section{Analysis of the attacks \label{results}}

Here we study the security thresholds $R=0$ that describe the performances of
the considered protocol for all possible attacks. The thresholds are given in
terms of the tolerable excess noise, defined as $N:=[T-1+(1-T)\omega]/T$,
as a function of the channel transmissivity $T$.

\begin{figure}[ptb]
\vspace{-0.0cm}
\par
\begin{center}
\includegraphics[width=0.35\textwidth]{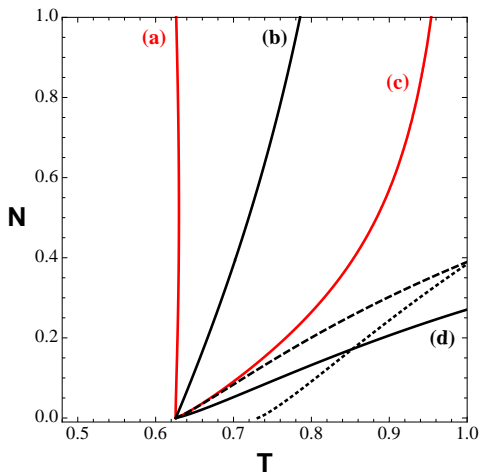}
\end{center}
\par
\vspace{-0.0cm}\caption{(Color online)Security thresholds for the case of
two-way protocol, in direct reconciliation, against two-mode coherent attacks.
In the ordinate is represented the excess noise (in vacuum shot noise units
(SNU)) and in the abscissa is represented the transmittivity. Curves $(a)$ and
$(c)$, describe two-mode attacks for which $g=-g^{\prime}$. In particular
$(a)$ is the threshold obtained when Eve uses maximally entangled ancillas
$E_{1}$ and $E_{2}$. This case is given by the two equivalent conditions on
the correlation parameters $g=\sqrt{\omega^{2}-1}=-$ $g^{\prime}$ and
$g=-\sqrt{\omega^{2}-1}=-$ $g^{\prime}$. Curve $(c)$ describes the cases
$g=\omega-1=-$ $g^{\prime}$ and $g=1-\omega=-$ $g^{\prime}$. The curves $(b)$
and $(d)$ correspond to the thresholds for $g=g^{\prime}$. In curve $(b)$ we
have $g=\omega-1=$ $g^{\prime}$and $g=1-\omega=g^{\prime}$ $(d)$. The dashed
line is the threshold for standard collective attacks, $g=g^{\prime}=0$. The
black dotted line is the security threshold for the corresponding one-way
protocol for which only collective attacks can be considered. We see that
curve $(d)$ partly goes below the one-way threshold for high
transmissivities.}%
\label{SUMMARY}%
\end{figure}

Fig. \ref{SUMMARY} shows the two-way security thresholds in direct
reconciliation. In particular the red lines, labeled by $(a)$ and $(c)$,
describe the thresholds of the two-way protocol obtained when the correlation
parameters of the attack fulfill the condition $g=-g^{\prime}$. In this case
curve $(a)$ describes the security threshold for maximally entangled ancillary
modes $E_{1}$ and $E_{2}$. This situation is described by two distinct
(despite equivalent) setup of the coherent attack, for which $|g|=\sqrt
{\omega^{2}-1}=-|g^{\prime}|$. Curve $(c)$, obtained when $|g|=\omega
-1=-|g^{\prime}|$, gives the extremal case of separable and maximally
correlated ancillas.
The black lines are the security thresholds when Eve exploits correlation of
the type $g=g^{\prime}$. In this group of attacks, modes $E_{1}$ and $E_{2}$
can only share separable correlation, and for $g=\omega-1=$ $g^{\prime}$ we
have curve $(b)$ while for $g=1-\omega=g^{\prime}$ we get curve $(d)$. Finally
the dashed line provides the two-way threshold, under standard collective
attacks, i.e., when $g=g^{\prime}=0$.

All these cases have been compared with the security threshold of the one-way
protocol \footnote{One-way means that in the figure we want to show not only
that the optimal 2-mode attack can reduce the security threshold of the
two-way protocol below that one against collective attacks, but even that for
large enough T, the optimal two-mode threshold is lower than the one-way
protocol against collective.}, in direct reconciliation (dotted line), for
which the collective attacks are known to be optimal. We see that
for\ standard collective attacks, the two-way protocol (dashed) always
overcome the performances of the one-way (dotted). However if Eve exploits
suitably correlated ancillas, she can perform a more profitable eavesdropping
of the two-way protocol. This is evident from curve $\left(  d\right)  $ which
is clearly below the security threshold corresponding to collective attacks
(dashed) and, for high transmissivity ($T\gtrsim0.86$), it goes below the
security threshold for the one-way protocol (dotted). Thus for the two-way
protocol described in this paper, we find that the two-mode coherent attack,
given by curve $(d),$ is optimal. In the Appendix we further deepen the
discussion about this result.

\section{Conclusions\label{conclusion}}

We have studied the two-way QKD protocol, focusing on its security under
two-mode coherent attacks. This represents the first study for this kind of
communication scheme in which a coherent attack can be explicitly considered
and analytically solved. The analysis spotlighted the first evidence of a
coherent attack beating the collective one in the setting of point-to-point protocols.

A similar result has been obtained in previous investigations focused on the
alternative approach to quantum cryptography, based on the end-to-end
paradigm. As proved in Refs.~\cite{RELAY,SIMMETRICO} when the parties
establish the key exploiting two channels with an untrusted middle relay, then
Eve can potentially obtain an advantage by exploiting correlated ancillary
modes. Here something similar happens, despite the optimal attack is different
\cite{NOTA-RELAY}.

Finally our analysis confirms the importance of the ON/OFF switching strategy,
in the context of two-way QKD\ protocols \cite{PIRS2}. In light of the results
presented, we conclude that the active exploitation of the additional degrees
of freedom available to the parties in two-way communication, represents a
necessary solution to avoid the possibility of powerful coherent attacks.
Alice can decide to open/close the two-way quantum communication, therefore
switching between one-way and two-way instances; finally Alice and Bob decide
which instances to keep on the base of Eve's strategy. In this sense the
ON/OFF switching can grant the immunity of two-way protocols against coherent
attacks. Further work \cite{2WAY-carlo} will extend these results here
restricted to direct reconciliation, and will consider finite-size effects and
composable security \cite{LEV1,LEV2}.

\acknowledgments The Authors acknowledge the financial support provided from
Leverhulme Trust and the EPSRC via `qDATA' (grant no. EP/L011298/1) and the
`UK Quantum Communications HUB' (Grant no. EP/M013472/1).

\appendix

\section{Optimal Attack\label{discussion}}

\begin{figure*}[t]
\vspace{-0.0cm}
\par
\begin{center}
\includegraphics[width=0.8\textwidth]{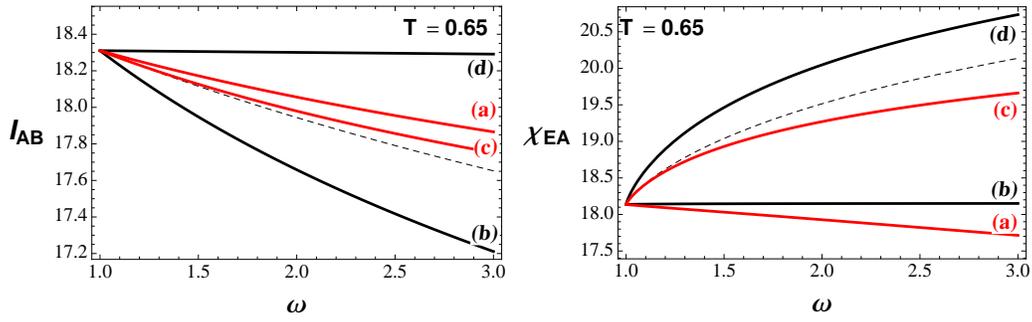}
\end{center}
\par
\vspace{-0.0cm}
\caption{(Color online)This figure shows the behavior of the asymptotic mutual
information $I_{AB}$ (left panel) and of the Holevo function $\chi_{EA}$
(right panel) as a function of Eve's thermal noise $\omega$. We fix the
Gaussian modulation $\mu=10^{6}$, value for which we checked the asymptotic
limit is achieved. We also fix the transmissivity $T=0.65$, for which the
parties may obtain a positive key-rate (see curves $(a)$ and $(b)$ in Fig.
\ref{SUMMARY}). The labeling corresponds to that adopted for the thresholds of
Fig. \ref{SUMMARY}. We have that $(a)$ describes two-mode attacks for which
$g=\sqrt{\omega^{2}-1}=-$ $g^{\prime}$, or $g=-\sqrt{\omega^{2}-1}=-$
$g^{\prime}$ and curve $(c)$ describes the cases $g=\omega-1=-$ $g^{\prime}$
or $g=1-\omega=-$ $g^{\prime}$. The curve $(b)$ corresponds to the case
$g=\omega-1=$ $g^{\prime}$and $(d)$ is for $g=1-\omega=g^{\prime}$, i.e., the
optimal attack. The dashed line refers to standard collective attacks,
$g=g^{\prime}=0$. We see that, for the optimal attack $(d)$, while the mutual
information slightly decreases by increasing $\omega$, the curve corresponding
to the Holevo bound, $\chi_{EA}$, increases and with a\ larger rate than in
any other attack. This causes the reduction of the key-rate in case $(d)$.}%
\label{muta-holevo-1}%
\end{figure*}

The result of Fig. \ref{SUMMARY} shows that differently from the one-way
protocol, the use of correlated ancillas is convenient for the eavesdropper.
To investigate further this feature we study the behavior of the quantities
defining the key-rate of Eq. (\ref{KEYRATE-AS}) as function of the thermal
noise $\omega$. We fix the classical Gaussian modulation $\mu=10^{6}$, for
which we have verified that the asymptotic limit is largely fulfilled, and the
transmissivity to the value $T=0.65$. In Fig. \ref{muta-holevo-1}, left panel,
we plot the mutual information $I_{AB}$, given in Eq. (\ref{IAB-ASY}), and in
the right panel we plot the Holevo function $\chi_{EA}$ given by Eq.
(\ref{CHI}).

\begin{figure*}[t]
\vspace{-0.0cm}
\par
\begin{center}
\includegraphics[width=0.8\textwidth]{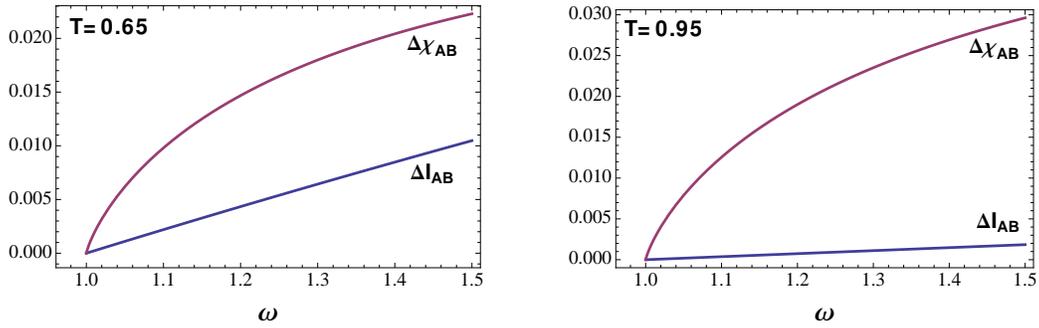}
\end{center}
\par
\vspace{-0.0cm}
\caption{(Color online)This figure shows the relative variation of the Holevo
bound $\Delta\chi_{EA}$, given in Eq.~(\ref{DELTACHIEA}) and of the mutual
information $\Delta I_{EA}$ of Eq.~(\ref{DELTAIAB}), for the optimal attack
(d), versus $\mathbf{\omega}$ for fixed values of the transmissivity, $T=0.65$
(left) and $T=0.95$ (right).}%
\label{MUTUAL-INFO-VS-HOLEVO-2}%
\end{figure*}

First, as one would expect, we note that the mutual information (left panel)
decreases for increasing thermal noise. Simultaneously Eve's accessible
information, $\chi_{EA}$ (right panel) corresponding to the optimal two-mode
attack $(d)$ is the highest among the others cases $(a)-(c)$. It also rapidly
increases for increasing $\omega$. This attack is profitable for Eve because
she is able to increase her knowledge on Alice's variable $\alpha$, at an
higher rate than Bob can do.
To illustrate further this property we have plotted in Fig.
\ref{MUTUAL-INFO-VS-HOLEVO-2} the relative variation of Alice-Bob mutual
information
\begin{equation}
\Delta I_{AB}=(I_{AB}-I_{c})/I_{c}, \label{DELTAIAB}%
\end{equation}
and of the Holevo function
\begin{equation}
\Delta\chi_{EA}=(\chi_{EA}-\chi_{c})/\chi_{c}, \label{DELTACHIEA}%
\end{equation}
of the optimal attack with respect to the respective expressions under
collective attacks ($g=g^{\prime}=0$), given by $I_{c}$ and $\chi_{c}$. In the
left panel we plot the case for $T=0.65$, while the right panel shows the case
$T=0.95$ We note that increasing the transmissivity $T$, the relative
variation in the mutual information tends to zero, while the relative
variation in Eve's Holevo information tends to increase.


\begin{thebibliography}{99}
\bibitem {NOTA-RELAY}In end-to-end QKD based on untrusted relays (performing a
CV Bell detection) the optimal attack, at fixed thermal noise $\omega$, is one
of the two of type $(a)$.
\end{thebibliography}
\end{document}